\documentclass[a4paper,12pt]{article}
\usepackage{amsmath, amsthm, amstext}                                                           %
\usepackage{amssymb,array, amsfonts}
\usepackage[dvips,colorlinks=true,linkcolor=blue,urlcolor=red]{hyperref}

\def \al{\alpha}
\def \be{\beta}
\def \ga{\gamma}
\def \de{\delta}
\def \er{\varepsilon}
\def \ze{\zeta}
\def \te{\theta}

\def \la{\lambda}
\def \si{\sigma}
\def \ph{\varphi}
\def \Ga{\Gamma}
\def \DD{\Delta}
\def \La{\Lambda}
\def \OO{\Omega}

\def \CC{\mathbb{C}}
\def \NN{\mathbb{N}}
\def \RR{\mathbb{R}}
\def \ZZ{\mathbb{Z}}
\def \C{\mathbb{C}}

\def \R{\mathbb{R}}
\def \Z{\mathbb{Z}}

\def\dom{\operatorname{Dom}}
\def\div{\operatorname{div}}
\def\mes{\operatorname{mes}}
\def\im{\operatorname{Im}}
\def\re{\operatorname{Re}}
\def\dd{\partial}
\def\n{\nabla}

\newcommand{\<}{\langle}
\renewcommand{\>}{\rangle}

\newtheorem{theorem}{Theorem}[section]
\newtheorem{remark}{Remark}[section]
\newtheorem{lemma}{Lemma}[section]

\title{Absolute continuity of the spectrum of a Schr{\"o}dinger operator
  with a potential which is periodic in some directions and decays in
  others}

\author{N. Filonov \footnote{N.F.'s research was partially supported
    by the FNS 2000 ``Programme Jeunes Chercheurs''.}\\
  Department of Mathematical Physics\\ St Petersburg State
  University\\ 198504 St Petersburg-Petrodvorets, Russia \\{\em
    email:}
  \href{mailto:filonov@mph.phys.spbu.ru}{filonov@mph.phys.spbu.ru}
  \and F. Klopp \footnote{F.K.'s research was partially supported by
    the program RIAC 160 at Universit{\'e} Paris 13 and by the FNS 2000
    ``Programme Jeunes Chercheurs''.}  \\
  LAGA, Institut Galil{\'e}e, Universit{\'e}
  Paris-Nord \\
  F-93430 Villetaneuse, France \\
  {\em email:}
  \href{mailto:klopp@math.univ-paris13.fr}{klopp@math.univ-paris13.fr}
}
\begin{document}
\maketitle
\section{Formulation of the result}
There are many papers (see, for example,~\cite{BSu,K}) devoted to the
question of the absolute continuity of the spectrum of differential
operators with coefficients periodic in the whole space. In the
present article, we consider the situation where the coefficients are
periodic in some variables and decay very fast (super-exponentially)
when the other variables tend to infinity. The corresponding operator
describes the scattering of waves on an infinite membrane or filament.
Recently, quite a few studies have been devoted to similar problems,
for periodic, quasi-periodic or random surface Hamiltonians (see,
e.g.~\cite{MR2001b:81026,MR2001m:47143,MR1968291}).\\
Let $(x,y)$ denote the points of the space $\mathbb{R}^{m+d}$. Define
$\OO=\RR^m\times (0, 2\pi)^d$ and $\< x \> = \sqrt{x^2+1}$. For $a\in\R$,
introduce the spaces
\begin{equation*}
  L_{p, a} = \{ f : e^{a \< x\>} f \in L_p (\OO)\}, \qquad H^2_a = \{
f : e^{a \< x\>} f \in H^2 (\OO)\},
\end{equation*}
where $1\le p\le \infty$ and $H^2(\OO)$ is the Sobolev space. Our main
result is
\begin{theorem}
  \label{main}
  Consider in $L_2 (\RR^{m+d})$ the self-adjoint operator
  \begin{equation}
    \label{1}
    H u = - \div (g \n u) + V u
  \end{equation}
  and assume that the functions $g:\ \R^{m+d}\to\R$ and $V:\ 
  \R^{m+d}\to\R$ satisfy following conditions:
  \begin{enumerate}
  \item $\forall l\in\ZZ^d$, $\forall(x,y)\in\R^{m+d}$,
    \begin{equation*}
      g (x,y+2\pi l) = g (x, y),\quad V(x,y+2\pi l) = V (x, y);
    \end{equation*}
  \item there exists $g_0>0$ such that $(g - g_0),\ \Delta g,\ V \in
    L_{\infty, a}$ for any $a>0$;
  \item there exists $c_0>0$ such that $\forall(x,y)\in\R^{m+d}$,
    $g(x,y)\geq c_0$.
  \end{enumerate}
  Then, the spectrum of $H$ is purely absolutely continuous.
\end{theorem}
\begin{remark}
  Operators with different values of $g_0$ differ from one another
  only by multiplication by a constant; so, without loss of
  generality, we can and, from now on, do assume that $g_0=1$.
\end{remark}
\begin{remark}
  If $V \equiv 0$,~\eqref{1} is the acoustic operator.  If $g\equiv
  1$, it is the Schr{\"o}\-din\-ger operator with electric potential $V$.
\end{remark}
The basic philosophy of our proof is the following. To prove the
absolute continuity of the spectrum for periodic operators (i.e.,
periodic with respect to a non degenerate lattice in $\R^d$), one
applies the Floquet-Bloch-Gelfand reduction to the operator and one is
left with proving that the Bloch-Floquet-Gelfand eigenvalues must vary
with the quasi-momentum i.e., that they cannot be constant on sets of
positive measure (see e.g.~\cite{K}). If one tries to follow the same
line in the case of operators that are only periodic with respect to a
sub-lattice, the problem one encounters is that, as the resolvent of
the Bloch-Floquet-Gelfand reduction of the operator is not compact,
its spectrum may contain continuous components and some
Bloch-Floquet-Gelfand eigenvalues may be embedded in these continuous
components. The perturbation theory of such embedded eigenvalues
(needed to control their behavior in the Bloch quasi-momentum) is more
complicated than that of isolated eigenvalues. To obtain a control on
these eigenvalues, we use an idea of the theory of resonances (see
e.g.~\cite{Zw}): if one analytically dilates Bloch-Floquet-Gelfand
reduction of the operator, these embedded eigenvalues become isolated
eigenvalues, and thus can be controlled in the usual way.
\par Let us now briefly sketch our proof.  We make the
Bloch-Floquet-Gelfand transformation with respect to the periodic
variables (see section 3) and get a family of operators $H(k)$ in the
cylinder $\OO$. Then, we consider the corresponding resolvent in
suitable weighted spaces. It analytically depends on the
quasi-momentum $k$ and the spectral (non real) parameter $\la$.  It
turns out that we can extend it analytically with respect to $\la$
from the upper half-plane to the lower one (see Theorem~\ref{RH}
below) and thus establish the limit absorption principle. This
suffices to prove the absolute continuity of the initial operator (see
section~\ref{sec:proof-theor-refm}).
\par Note that an analytic extension of the resolvent of the operator
(\ref{1}) with coefficients $g$ and $V$ which decay in all directions
is constructed in the paper~\cite{E} (with $m = 3$, $d = 0$; see also
\cite{La} for $g \equiv 1$). In the case of a potential decaying in
all directions but one (i.e., if $d = 1$), the analytic extension of
the resolvent of the whole operator (\ref{1}) (not only for the
operator $H(k)$ (see section~\ref{sec:resolv-free-oper})) is
investigated in~\cite{Ge} when $g \equiv 1$.  Note also that our
approach has shown to be useful in the investigation of the
perturbation of free operator in the half-plane by $\de$-like
potential concentrated on a line (see \cite{FrSh}); the wave operators
are also constructed there.
\par In section~\ref{sec:auxil-estim}, we establish some auxiliary inequalities. In
section~\ref{sec:resolv-free-oper}, we define the Floquet-Gelfand
transformation and construct an analytic extension of the resolvent of
free operator in the cylinder $\OO$. In
sections~\ref{sec:invert-oper-type}
and~\ref{sec:resolvent-operator-h}, we prove a limiting absorption
principle for the initial operator in the cylinder. An auxiliary fact
from theory of functions is established in
section~\ref{sec:one-fact-from}. Finally, the proof of
Theorem~\ref{main} is completed in section~\ref{sec:proof-theor-refm}.
\vskip.2cm\noindent We denote by $B_\de (k_0)$ a ball in real space
\begin{equation*}
  B_\de (k_0) = \{ k \in \RR^d : |k - k_0| < \de \}
\end{equation*}
and by $k_1$ the first coordinate of $k$, $k = (k_1, k')$.  We will
use the spaces of function in $\OO$ with periodic boundary conditions,
\begin{gather*}
  \tilde H^2 = \left\{ f \in H^2 (\OO) : f\mid_{y_i = 0} = f\mid_{y_i
      = 2\pi}, \ \frac{\dd f}{\dd y_i}\mid_{y_i = 0} = \frac{\dd
      f}{\dd y_i}\mid_{y_i = 2\pi}, \ 
    i = 1, \dots, d \right\},\\
  \tilde H^2_{loc} =\left\{ f \in H^2_{loc} (\OO) : f\mid_{y_i = 0}=
    f\mid_{y_i = 2\pi}, \ \frac{\dd f}{\dd y_i}\mid_{y_i = 0} =
    \frac{\dd f}{\dd y_i}\mid_{y_i = 2\pi}, \ i = 1, \dots, d
  \right\}.
\end{gather*}
Finally $B(X,Y)$ is the space of all bounded operators from $X$ to
$Y$, and $B(X)=B(X,X)$, both endowed with their natural topology.

\noindent{\it Thanks:} the authors are grateful to Prof. P.~Kuchment
for drawing their attention to the question addressed in the present
paper, and to Prof.  T.~Suslina for useful discussions.

\section{Auxiliary estimations}
\label{sec:auxil-estim}

In this section, we assume that the pair $(k_0, \la_0) \in \RR^{d+1}$
satisfies
\begin{equation}
  \label{2}
  (k_0 + n)^2 \not= \la_0 \qquad \forall n \in \ZZ^d.
\end{equation}
The constants in all the inequalities in this section may depend on
$(k_0, \la_0)$. The set
\begin{equation}
  \label{eq:4}  
  J = \{ n \in \ZZ^d : (k_0 + n)^2 < \la_0 \}
\end{equation}
is finite. In a neighborhood of $(k_0, \la_0)$, the partition of
$\ZZ^d$ into $J$ and $(\ZZ^d \setminus J)$ is clearly the same.  In
other words, there exists $\de = \de (k_0, \la_0) > 0$ such that
\begin{equation}
  \label{3}
  \text{if } k \in B_\de (k_0)\text{ and }\la\in B_\de (\la_0),
  \text{ then }\ (k + n)^2 < \la \Leftrightarrow n \in J.
\end{equation}
Choose $\tilde k \in B_\de (k_0)$ with $\tilde k_1 \notin \ZZ$ and put
\begin{gather}
  k (\tau) := (\tilde k_1 + i \tau, \tilde k') \in \CC^d, \quad \tau
  \in\RR,\nonumber\\
  \intertext{and}
  \label{5}
  M_1 = M_1 (k_0, \la_0) := \left( B_\de (k_0) \cup \{ k(\tau) \}_{\tau
      \in \RR} \right) \times B_\de (\la_0).
\end{gather}
\begin{lemma}
  \label{2.1}
  There exists $c>0$ such that, for all $\ze\in\RR^m$, $(k,\la)\in
  M_1$, $n\in\ZZ^d \setminus J$ and $\tau \in \RR$, we have
  \begin{gather*}
    |\ze^2 + (k + n)^2 - \la|\geq c,\\
    |\ze^2 + (k (\tau) + n)^2 - \la| \ge c|\tau|.
  \end{gather*}
\end{lemma}

\begin{proof}
  By virtue of (\ref{3}), there exists $c>0$ such that, for
  $n\in\Z\setminus J$, 
  \begin{equation*}
    \forall k \in B_\de (k_0),\
    \forall \lambda\in B_\de (\lambda_0),\quad (k + n)^2 - \la > c. 
  \end{equation*}
  Hence, for $\zeta\in\R^m$, $n\in\Z\setminus J$, 
  \begin{equation*}
    \forall k \in B_\de (k_0),\ \forall \lambda\in B_\de
    (\lambda_0),\quad \ze^2 + (k + n)^2 - \la >c.
  \end{equation*}
  The second inequality is an immediate corollary of our choice of
  $\tilde k_1$ and the equality
  \begin{equation*}
    \im (\ze^2 + (k (\tau) + n)^2 - \la) = 2 (\tilde k_1 + n_1)
    \tau.
  \end{equation*}
  This completes the proof of Lemma~\ref{2.1}.
\end{proof}

\noindent In the remaining part of this section, we assume
$\la_0>0$. In this case, we will need to change the integration path
in the Fourier transformation; we now describe the contour
deformation. Fix $\eta>\sqrt\la_0$ and, let $\gamma$ be the contour in
the complex plane defined as
\begin{equation}
  \label{6}
  \ga = \{ -\xi + i \eta \}_{\xi\in [\eta,\infty)}
  \cup \{ \al (1-i) \}_{\al\in [-\eta,\eta]}
  \cup \{ \xi - i \eta \}_{\xi\in [\eta,\infty)}.
\end{equation}
Two following assertions are clear.
\begin{lemma}
  \label{2.2}
  If $g \in L_2(\ga)$ and $\eta_0 > \eta$ then the function
  \begin{equation*}
    h(t) = e^{-\eta_0 |t|} \int_\ga e^{itz} g(z) dz
  \end{equation*}
  belongs to $L_2(\RR)$.
\end{lemma}
\begin{lemma}
  \label{2.3}
  Let $\Ga$ denote the open set between real axis and $\ga$ (it
  consists of two connected components). Let $g$ be an analytic
  function in $\Ga$ such that $g\in C(\overline\Ga)$ and $|g(z)|\leq C
  (1+|\re z|)^{-2}$.  Then,
  \begin{equation*}
    \int_\RR e^{itz} g(z) dz = \int_\ga e^{itz} g(z) dz
    \qquad \forall t \in \RR.
  \end{equation*}
\end{lemma}
\noindent Establish an analogue of Lemma~\ref{2.1} for $n\in J$ and
$\zeta\in\gamma^m$ i.e., $\ze=(\zeta_1,\dots,\zeta_m)\in\CC^m$,
$\ze_j\in\ga$.
\begin{lemma}
  \label{2.4}
  Let $\la_0 > 0$, $\eta > \sqrt\la_0$ and $\ga$ be defined by
  (\ref{6}). There exists $c>0$ such that, for all $\ze\in\ga^m$,
  $(k,\la)\in M_1$, $n \in J$ and $\tau \in \RR$, we have
  \begin{gather}
    |\ze^2 + (k + n)^2 - \la|\geq c,\nonumber\\
    \label{eq:7}
    |\ze^2 + (k (\tau) + n)^2 - \la| \ge c|\tau|.
  \end{gather}
\end{lemma}

\begin{proof}
  By virtue of~\eqref{3}, there exists $\tilde c>0$ such that, for
  $n\in J$,
  \begin{equation*}
    \forall k \in B_\de (k_0),\
    \forall \lambda\in B_\de (\lambda_0),\quad (k + n)^2 - \la <-2\tilde c. 
  \end{equation*}
  Hence, for $\zeta\in\gamma^m$ such that $|\zeta|\leq \sqrt{\tilde
    c}$, one has
  \begin{equation*}
    \forall k \in B_\de (k_0),\
    \forall \lambda\in B_\de(\lambda_0),\quad \re(\zeta^2+(k + n)^2 -
    \la) <-\tilde c.
  \end{equation*}
  On the other hand, for $\zeta\in\gamma^m$ such that
  $|\zeta|\geq\sqrt{\tilde c}$, one has
  \begin{equation*}
    \forall k \in B_\de (k_0),\
    \forall \lambda\in B_\de(\lambda_0),\quad \im(\zeta^2+(k + n)^2 -
    \la) <-\tilde c
  \end{equation*}
  if one chooses $\tilde c$ sufficiently small. Thus, it remains to
  prove the second inequality. Therefore, we write
  \begin{equation*}
    \ze^2 = - 2i \sum_p \al_p^2 + \sum_q (\xi_q - i\eta)^2,
  \end{equation*}
  where the indexes $p$ correspond to the coordinates of $\ze$ which
  are in the middle part of $\ga$ (i.e., $|\re\zeta_p|<\eta$) and the
  indexes $q$ correspond to the extreme parts of $\ga$ (i.e.,
  $|\re\zeta_q|\geq\eta$); it is possible that there are only indexes
  $p$ or only $q$.  Without loss of generality, we suppose that, for
  all $q$, $\xi_q \ge 0$. Thus,
  \begin{equation*}
    \begin{split}
      \ze^2+(k(\tau)+n)^2-\la&=
      \sum_q(\xi_q^2-\eta^2)+(\tilde k+n)^2-\tau^2-\la\\
      &\hskip2cm+ 2i \left( - \sum_p \al_p^2 - \sum_q \xi_q \eta +
        (\tilde k_1 + n_1) \tau \right).\nonumber
    \end{split}
  \end{equation*}
  Fix some $\si \in (\eta^{-1} \sqrt\la_0, 1)$.  If $\sum_q \xi_q \ge
  \si |\tau|$ then,
  \begin{equation*}
    \left|\im (\ze^2 + (k (\tau) + n)^2 - \la)\right| \ge 2 \left(\si
      \eta - |\tilde k_1 + n_1|\right) |\tau| > 2(\si \eta - \sqrt\la_0)
    |\tau|,
  \end{equation*}
  as $(\tilde k + n)^2 < \la_0$.  If $\sum_q \xi_q \le \si |\tau|$
  then $\sum_q \xi_q^2 \le \si^2 \tau^2$ and
  \begin{equation*}
    \left|\re (\ze^2 + (k (\tau) + n)^2 - \la)\right| \ge
    \tau^2 + \la - (\tilde k + n)^2 - \si^2 \tau^2
    > (1-\si^2) \tau^2
  \end{equation*}
  again by virtue of (\ref{3}). This completes the proof of
  Lemma~\ref{2.4}.
\end{proof}

\section{The resolvent of free operator in the cylinder}
\label{sec:resolv-free-oper}

Let us consider the Floquet-Gelfand transformation
\begin{equation*}
  (Uf)(k, x, y) = \sum_{l\in\ZZ^d}e^{i\<k,y+2\pi l\>}f(x,y+2\pi l).
\end{equation*}
It is a unitary operator
\begin{equation*}
  U : L_2 (\RR^{m+d}) \to \int_{[0, 1)^d}^\oplus L_2 (\OO) dk.
\end{equation*}
Introduce the family of operators $(H(k))_{k\in\C^d}$ on the cylinder
$\OO$ where for $k\in\C^d$, $\dom H(k) = \tilde H^2$ and
\begin{equation}
  \label{7}
  H(k) = \left( i \n - (0, \overline k)\right)^*
  g(x, y) \left( i \n - (0, k)\right) + V(x, y).
\end{equation}
Then, the Schr{\"o}dinger operator (\ref{1}) is unitarily equivalent to
the direct integral of these operators in $\OO$:
\begin{equation*}
  U H U^* = \int_{[0, 1)^d}^\oplus H (k) dk.
\end{equation*}

\noindent In this section, we investigate the free operator
\begin{equation}
  \label{9}
  A(k) = - \DD_x + \left( i \n_y - \overline k \right)^*
  \left( i \n_y - k \right)
\end{equation}
(which corresponds $H(k)$ with $g\equiv 1$, $V\equiv0$). For
$k\in\RR^d$ and $\la\not\in\R$, its resolvent can be expressed as
\begin{equation}
  \label{10}
  \left( (A(k) - \la)^{-1} f\right) (x, y) =
  \sum_{n\in\ZZ^d} \int_{\RR^m} \frac{e^{i\ze x + iny} (Ff) (\ze, n) d\ze}
  {\ze^2 + (k+n)^2 - \la},
\end{equation}
where $F$ denotes the Fourier transformation in the cylinder
\begin{equation*}
  (Ff) (\ze, n) = (2\pi)^{-m-d}
  \int_\OO e^{-i\ze x - iny} f(x, y)\,dx\,dy.
\end{equation*}
Let $(k_0,\la_0)\in\RR^{d+1}$ satisfy (\ref{2}) and, $J$ and $M_1$ be
defined respectively by formulas (\ref{eq:4}) and (\ref{5}) in the
previous section.
\begin{lemma}
  \label{3.1}
  There exists $\mathcal{V}_1$, a neighborhood of the set $M_1$ in
  $\CC^{d+1}$ such that, for $(k,\lambda)\in\mathcal{V}_1$, the
  operator $R_1 (k, \la)$ given by
  \begin{equation*}
    \left( R_1 (k, \la) f\right) (x, y) = \sum_{n\in\ZZ^d\setminus J}
    \int_{\RR^m} \frac{e^{i\ze x + iny} (Ff) (\ze, n) d\ze} {\ze^2 +
      (k+n)^2 - \la}
  \end{equation*}
  is well defined and is bounded from $L_2 (\OO)$ to $H^2 (\OO)$. The
  $B (L_2(\OO), H^2(\OO))$-valued function $(k,\lambda)\mapsto
  R_1(k,\lambda)$ is analytic in $\mathcal{V}_1$. For $\tau\not=0$,
  the estimate
  \begin{equation*}
    \| R_1 (k(\tau), \la)\|_{B(L_2(\OO))} \le C |\tau|^{-1}  
  \end{equation*}
  holds.
\end{lemma}
\begin{proof}
  It immediately follows from Lemma \ref{2.1}.
\end{proof}

\begin{lemma}
  \label{3.2}
  Let $\la_0 > 0$, $\eta > \sqrt\la_0$, $a > \eta\sqrt m$ and the
  contour $\ga$ be defined by (\ref{6}). Then, there exists a
  neighborhood of the set $M_1$, say $\mathcal{V}_2$, such that, for
  $(k,\lambda)\in\mathcal{V}_2$, the operator $R_2 (k,\la)$ given by
  \begin{equation}
    \label{11}
    \left( R_2 (k, \la) f\right) (x, y) =
    \sum_{n\in J}
    \int_\ga \cdots \int_\ga \frac{e^{i\ze x + iny} (Ff) (\ze, n)}
    {\ze^2 + (k+n)^2 - \la}\,d\ze_1 \cdots d\ze_m
  \end{equation}
  is well defined as a bounded operator from $L_{2, a}$ to $H^2_{-a}$.
  The $B(L_{2,a},H^2_{-a})$-valued function $(k,\lambda)\mapsto
  R_2(k,\lambda)$ is analytic in $\mathcal{V}_2$. For $\tau\not=0$,
  the estimate
  \begin{equation*}
    \| R_2 (k(\tau), \la)\|_{B(L_{2,a},\,L_{2,-a})} \le C |\tau|^{-1}  
  \end{equation*}
  holds.
\end{lemma}

\begin{proof}
  If $f\in L_{2, a}$ then the function $(Ff) (\cdot, n)$ is square
  integrable on $\ga^m$. By Lemma \ref{2.4}, the denominator in
  (\ref{11}) never vanishes for $(k, \la) \in M_1$; therefore, in some
  neighborhood of $M_1$. So
  \begin{equation*}
    \left| (\ze^2 + (k+n)^2 - \la)^{-1} e^{i\ze x + iny}\right| \le C
    |e^{i\ze x}|
  \end{equation*}
  where the constant does not depend on $\ze\in\ga^m$ and on $x$; the
  same is true for the second derivatives of $(\ze^2+(k+n)^2-\la)^{-1}
  e^{i\ze x+iny}$ with respect to $(x,y)$. Hence, $R_2(k,\la)\in B
  (L_{2,a},H^2_{-a})$ by virtue of Lemma \ref{2.2}.
  Estimation~\eqref{eq:7} yields the estimation for the norm of $R_2
  (k(\tau),\la)$.
\end{proof}
\noindent Now, we construct an analytic extension of the resolvent of
  $A(k)$.
\begin{theorem}
  \label{freeres}
  Let $(k_0, \la_0)\in \RR^{d+1}$ satisfy (\ref{2}) and the set $M_1$
  be defined in~(\ref{5}). Then, there exists a neighborhood of
  $M_1$in $\CC^{d+1}$, say $M_0$, a real number $a$ and a
  $B(L_{2,a},H^2_{-a})$-valued function, say $(k,\la)\mapsto
  R_A(k,\la)$, defined and analytic in $M_0$, such that, for $(k, \la)
  \in M_0$, $k\in \RR^d$, $\im \la > 0$ and $f \in L_{2, a}$, one has
  \begin{equation}
    \label{12}
    R_A (k, \la) f = (A(k) - \la)^{-1} f
  \end{equation}
  and
  \begin{equation}
    \label{13}
    \| R_A (k(\tau), \la)\|_{B(L_{2,a},\,L_{2,-a})} \le C |\tau|^{-1}.
  \end{equation}
\end{theorem}
\begin{proof}
  If $\la_0 \le 0$, we can take $R_A = R_1$ ($R_1$ is constructed in
  Lemma \ref{3.1}; here, $J = \emptyset$ and $a=0$).
  
  If $\la_0 > 0$ then, we put $R_A = R_1 + R_2$, where $R_1$, $R_2$
  and $a$ are defined in Lemmas \ref{3.1} and \ref{3.2}, and $M_0$ is
  the intersection of $\mathcal{V}_1$ and $\mathcal{V}_2$defined
  respectively in Lemma~\ref{3.1} and Lemma~\ref{3.2}. If $f\in L_{2,
    a}$ then $(Ff) (\cdot, n)$ is an analytic function in the domain
  $\{ \ze : |\im\ze| < a \}$ and is uniformly bounded on $\{ \ze :
  |\im\ze| \le \eta \sqrt m \}$.  If $\ze \in \overline\Ga^m$ where
  $\Ga$ is the open set between $\RR$ and $\ga$ (see Lemma \ref{2.3}),
  then, $\im \ze^2 \le 0$; therefore, the integrand in (\ref{11}) has
  no poles when $\im\la > 0$. Hence, the integral in right hand side
  of (\ref{10}) for $n\in J$ coincides with the corresponding integral
  in (\ref{11}) due to Lemma \ref{2.3}, and (\ref{12}) holds.

  The estimate (\ref{13}) is a simple corollary of the estimations of
  Lemmas \ref{3.1} and \ref{3.2}.
\end{proof}

\section{Invertibility of operators of type $(I + W R_A)$}
\label{sec:invert-oper-type}

\begin{lemma}
\label{4.1}
Let $W \in L_{\infty, b}$ for $b > 2a > 0$.  Then, the operator of
multiplication by $W$ (we will denote it by the same letter) is
\begin{enumerate}
\item bounded as an operator from $L_{2, -a}$ to $L_{2, a}$;
\item compact as an operator from $H^2_{-a}$ to $L_{2, a}$.
\end{enumerate}
\end{lemma}

\begin{proof}
  The first assertion is evident.  In order to prove the second it is
  enough to introduce functions
  \begin{equation*}
    W_\rho (x, y) =\begin{cases} W (x, y),& |x| < \rho, \\
      0,& |x|\geq\rho, \\ \end{cases}
  \end{equation*}
  and note that the multiplication by $W_\rho$ is a compact operator
  from $H^2_{-a}$ to $L_{2,a}$ and that
  \begin{equation*}
    \| W - W_\rho\|_{B(L_{2, -a}, L_{2, a})} \to 0
  \end{equation*}
  when $\rho \to \infty$.
\end{proof}
\noindent The next lemma is a well known result from analytic Fredholm
theory (see, e.g.,~\cite{Ka:80,Re-Si:80}).
\begin{lemma}
  \label{4.2}
  Let $U$ be a domain in $\CC^p$, $z_0 \in U$.  Let $z\mapsto T(z)$ be
  an analytic function with values in the set of compact operators in
  some Hilbert space $\cal H$. Then, there exists a neighborhood $U_0$
  of the point $z_0$ and an analytic function $h:\ U_0\to\C$ such
  that, for $z\in U_0$,
  \begin{equation*}
    \left(I+T(z)\right)^{-1}\text{ exists if and only if }h(z)=0.  
  \end{equation*}
\end{lemma}
\noindent Now, we can establish the existence of the inverse of
$(I+WR_A)$.
\begin{theorem}
  \label{inv}
  Let $(k_0,\la_0)$ satisfy (\ref{2}), $R_A (k, \la)$ and $a$ be
  defined as in Theo\-rem~\ref{freeres}. Pick $b>2a$, and let
  $(x,y,\la) \mapsto W(x,y,\la)$ be a function which belongs to
  $L_{\infty,b}$ for all $\la$, and is analytic with respect to $\la$
  i.e., $\lambda\mapsto W(\cdot,\cdot,\lambda)\in
  Hol(\CC,L_{\infty,b})$.\\
  Then, there exists $\er>0$, an open set $U\subset\CC^{d+1}$ such
  that $B_\er(k_0)\times B_\er(\la_0)\subset U$, and an analytic function
  $h:\ U\to\C$ such that
  \begin{equation}
    \label{14}
    \forall\la\in B_\er(\la_0),\quad\exists k\in B_\er(k_0)\quad
    \text{such that}\quad h(k,\la)\not= 0,
  \end{equation}
  and, for any $(k,\la)\in U$, the operator $(I+W(\la)R_A (k,\la))$ is
  invertible in $L_{2,a}$ if and only if $h(k,\la)\not=0$.
\end{theorem}
\begin{proof}
  Due to Theorem \ref{freeres} and Lemma \ref{4.1}, the operator
  $W(\la) R_A (k(\tau), \la)$ is compact in $L_{2, a}$ and satisfies
  the inequality
  \begin{equation*}
    \|W(\la)R_A(k(\tau),\la)\|_{B(L_{2,a})}\le C|\tau|^{-1}, \quad
    \forall\la\in B_\er(\la_0).
  \end{equation*}
  Therefore, for $|\tau|$ large enough, the operator $(I+W(\la)R_A
  (k(\tau),\la))^{-1}$ exists and is bounded on $L_{2,a}$. The
  operator-valued function $\lambda\mapsto W(\la)R_A(k,\la)$ is
  analytic in $M_0$ (defined in Theorem~\ref{freeres}). The analytic
  Fredholm alternative yields that, for each $\la\in B_\er(\la_0)$,
  one can find $k\in B_\er(k_0)$ such that the operator
  $(I+W(\la)R_A(k,\la))^{-1}$ exists. Now, applying Lemma~\ref{4.2}
  with ${\cal H}=L_{2, a}$, $z=(k,\la)$ and $T(z)=WR_A$, completes the
  proof of Theorem~\ref{inv}.
\end{proof}
\section{The resolvent of the operator $H$}
\label{sec:resolvent-operator-h}
We can reduce the general case of operator (\ref{1}) with a ``metric''
$g$ to the case of ``pure'' Schr{\"o}dinger operator due to the following
lemma.  This identity (for the totally periodic case) is known
(see~\cite{BSu}). We include the proof for the convenience of the
reader.
\begin{lemma}
  \label{5.1}
  Let the operators $H(k)$ and $A(k)$ be defined by (\ref{7}) and
  (\ref{9}) respectively, and let the conditions of Theorem~\ref{main}
  be fulfilled with $g_0 = 1$. If $u \in \tilde H^2$ then,
  \begin{equation*}
    \left( H(k) - \la\right) g^{-1/2} u =
    g^{1/2} \left( A(k) + W(\la) - \la\right) u,
  \end{equation*}
  where
  \begin{equation}
    \label{15}
    W(\la) = \frac{1}{g} \left( \frac{\DD g}{2} -
      \frac{|\n g|^2}{4 g} + V + \la (g - 1) \right).
  \end{equation}
\end{lemma}
\begin{remark}
If $g\equiv 1$ then $W(\la)\equiv V$.
\end{remark}
\begin{proof}
  It is enough to prove the equality
  \begin{equation}
    \label{16}
    \left(i\n - (0, \overline k)\right)^* g
    \left(i\n - (0, k)\right) (g^{-1/2} u) =
    g^{1/2} \left(A(k) + \frac{\DD g}{2g} - \frac{|\n g|^2}{4 g^2}\right) u.
  \end{equation}
  We have
  \begin{equation*}
    \left(i\n - (0, k)\right) (g^{-1/2} u) =
    i g^{-1/2} \n u - \frac{i}{2} g^{-3/2} \n g u - (0, k) (g^{-1/2} u).  
  \end{equation*}
  Therefore, the left hand side of (\ref{16}) is equal to
  \begin{equation*}
    \begin{split}
      (i\n&-(0,\overline k))^*\left(ig^{1/2}\n
        u-\frac{i}{2}g^{-1/2}\n gu-(0,k)(g^{1/2}u)\right)\\
      &=-g^{1/2}\DD u+\frac{1}{2}\div(g^{-1/2}\n g)u
      -i\<k,\n_y(g^{1/2}\overline u)\>_\CC\\
      &\hskip3cm -ig^{1/2}\<\n_y u,\overline k\>_\CC
      +\frac{i}{2}g^{-1/2} \<\n_y g,\overline k\>_\CC u+k^2g^{1/2}u\\
      &=g^{1/2}\left(-\DD_x u+(i\n_y-\overline k)^*(i\n_y-k)
        u+\frac{1}{2}g^{-1/2}\div(g^{-1/2}\n g)u\right).
    \end{split}
  \end{equation*}
  This completes the proof of Lemma~\ref{5.1}.
\end{proof}
\noindent In the following theorem, we describe the meromorphic
extension of the resolvent of $H(k)$.
\begin{theorem}
  \label{RH}
  Let the conditions of Theorem~\ref{main} be fulfilled, the operator
  $H(k)$ be defined by (\ref{7}) and $(k_0, \la_0) \in \RR^{d+1}$
  satisfy (\ref{2}). Then, there exists numbers $a\ge 0$, $\er >0$, a
  neighborhood $U$ of $(k_0, \la_0)$ in $\CC^{d+1}$ containing the set
  $B_\er (k_0) \times B_\er (\la_0)$, a function $h \in Hol (U)$ satisfying
  (\ref{14}) and an operator-valued function $(k,\la)\mapsto
  R_H(k,\la)$ having the following properties:
  \begin{enumerate}
  \item $R_H$ is defined on the set $\{(k,\la)\in U: h(k,\la)\not=0\}$
    and is analytic there;
  \item for $(k,\la)\in U$ such that $h(k,\la)\not=0$, one has $R_H (k,
    \la) \in B (L_{2, a}, L_{2, -a})$;
  \item for $(k, \la) \in U$, $k\in \RR^d$, $\im \la > 0$, $f \in L_{2, a}$
    \begin{equation}
      \label{17}
      R_H (k, \la) f = (H(k) - \la)^{-1} f.
    \end{equation}
  \end{enumerate}
\end{theorem}
\begin{remark}
  It will be seen from the proof that $R_H (k, \la) \in B (L_{2, a},
  H^2_{-a})$ though we do not need this fact.
\end{remark}

\begin{proof}
  By the assumptions of Theorem~\ref{main}, for any $b>0$, $\n g\in
  L_{\infty,b}$. So, if we define $W(\la)$ by (\ref{15}), for any
  $b>0$, $W(\la)\in L_{\infty,b}$. We can thus apply
  Theorem~\ref{inv}.  Let $U$, $h$, $a$ and $R_A$ be as in this
  theorem. On the set where $h(k,\la)\not=0$, we put
  \begin{equation*}
    R_H (k, \la) = g^{-1/2} R_A (k, \la) \left( I + W(\la) R_A (k,
      \la)\right)^{-1} g^{-1/2}.
  \end{equation*}
  By Theorem~\ref{4.1}, $R_H (k, \la) \in B (L_{2, a}, H^2_{-a})$.
  Let $f \in L_{2, a}$.  Then,
  \begin{equation}
    \label{18}
    \left( I + W(\la) R_A (k, \la)\right)^{-1} g^{-1/2} f \in L_{2, a}
  \end{equation}
  and we can apply Lemma~\ref{5.1} to the function
  \begin{equation}
    \label{19}
    u = R_A (k, \la) \left( I + W(\la) R_A (k, \la)\right)^{-1}
    g^{-1/2} f \in H^2_{-a},
  \end{equation}
  so
  \begin{equation}
    \label{20}
    (H(k) - \la) R_H (k, \la) f =
    g^{1/2} \left( A(k) + W(\la) - \la\right) u.
  \end{equation}
  For real $k$ and non real $\la$, we have by (\ref{12}) and
  (\ref{18})
  \begin{equation*}
    (A(k) - \la) u =
    \left( I + W(\la) R_A (k, \la)\right)^{-1} g^{-1/2} f,  
  \end{equation*}
  hence, by (\ref{19}),
  \begin{equation*}
    \left( A(k) + W(\la) - \la\right) u = g^{-1/2} f,  
  \end{equation*}
  and, finally, by (\ref{20})
  \begin{equation}
    \label{eq:6}
    (H(k) - \la) R_H (k, \la) f = f.  
  \end{equation}
  For $\im\la>0$, the operators $(H(k)-\la)^{-1}$ and
  $(A(k)-\la)^{-1}$ are well defined in $L_2(\OO)$. As $R_H (k,\la)f
  \in L_2(\OO)$,~\eqref{eq:6} gives $R_H(k,\la)f=(H(k)-\la)^{-1}f$.
  This completes the proof of Theorem~\ref{RH}.
\end{proof}
\section{One fact from the theory of functions}
\label{sec:one-fact-from}
\begin{lemma}
  \label{6.1}
  Let $U$ be an open subset of $\RR^d$. 
  Let $f$ be a real-analytic function on the set $U\times (a, b)$, 
  and pick $\La \subset (a, b)$ such that $\mes \La = 0$. 
  Then
  \begin{equation}
    \label{eq:3}
    \mes\{k\in U :\ \exists\la\in\La\text{ s.t. }f(k,\la)=0
    \text{ and }\dd_{k_1}f(k,\la)\ne0\}=0.
  \end{equation}
\end{lemma}

\begin{proof}
  The Implicit Function Theorem implies that, for any point 
  $(k^*,  \la^*)$ such that 
  $f(k^*, \la^*) = 0 \ne \dd_{k_1} f(k\*, \la^*)$, 
  we can find rational numbers $\tilde r>0$, $\tilde\la$,
  a vector $\tilde k$ with rational coordinates, 
  and a real analytic function $\te$ defined in 
  $B_{\tilde r} (\tilde k',\tilde\lambda)$ such that
  \begin{enumerate}
  \item $(k^*,  \la^*) \in B_{\tilde r} (\tilde k,\tilde\lambda)$;
  \item $\te ((k^*)', \la\*) = k_1^*$;
  \item $f(k, \la) = 0 \Leftrightarrow \te(k',\la) = k_1$ if 
  $(k, \la) \in B_{\tilde r}(\tilde k,\tilde\lambda)$.
  \end{enumerate}
  The Jacobian of the map
  \begin{equation*}
    (k', \la) \mapsto (\te(k', \la), k')
  \end{equation*}
  is bounded, so
  \begin{equation*}
    \mes\{ (\te(k', \la), k') : (k', \la) 
  \in B_{\tilde r} (\tilde k',\tilde\lambda), \la \in \La\} = 0,
  \end{equation*}
  and therefore,
  \begin{equation*}
    \mes\{ k : \exists\la\in\La\text{ s.t. }(k, \la) 
   \in B_{\tilde r}(\tilde k,\tilde\lambda)\text{ and } f(k,\la)=0\} =0.
  \end{equation*}
  The set
  \begin{equation*}
    \{ (k, \la) : f (k, \la) = 0\text{ and }\dd_{k_1}f(k, \la) \ne 0 \}  
  \end{equation*}
  can be covered by a countable number of balls 
  $B_{\tilde r_i}(\tilde k_i,\tilde\lambda_i)$ 
  constructed as above, hence the measure of the set
  in~\eqref{eq:3} is also equal to zero.
\end{proof}
\begin{theorem}
  \label{func}
  Let $U$ be a region in $\RR^d$, $\La$ be a subset of an interval
  $(a, b)$ such that $\mes \La = 0$. Let $h$ be a real-analytic
  function defined on the set $U\times (a,b)$ and suppose that
  \begin{equation}
    \label{eq:8}
    \forall \la \in \La \quad \exists k \in U \quad
    \text{such that} \quad h(k, \la) \ne 0.
  \end{equation}
  Then,
  \begin{equation*}
    \mes\{k\in U :\ \exists\la\in\La\text{ s.t. }h(k,\la)=0\}=0.  
  \end{equation*}
\end{theorem}
\begin{proof}
  For any $k \in U$ and $\la \in \La$, by assumption~\eqref{eq:8},
  there exists a multi-index $\alpha \in \ZZ_+^d$ such that
  $\partial_k^\alpha h (k, \la) \ne 0$.  Therefore,
  \begin{align}
    \{ k \in U &: h (k, \la) = 0
    \text{ for some } \la \in \La \} \nonumber\\
    \subset &\bigcup_{j=1}^d \bigcup_{\alpha\in \ZZ_+^d}
    \left\{ k \in U : \partial_k^\alpha h (k, \la) = 0,
      \dd_{k_j} \partial_k^\alpha h (k, \la) \ne 0
      \text{ for some } \la \in \La \right\}.\nonumber
  \end{align}
  Reference to Lemma \ref{6.1} then completes the proof of
  Theorem~\ref{func}.
\end{proof}
\section{The proof of Theorem \ref{main}}
\label{sec:proof-theor-refm}
The following lemma is well known (see for example \cite{Ya}).
\begin{lemma}
  \label{7.1}
  Fix $b>0$. Let $B$ be a self-adjoint operator in $L_2(\OO)$. Suppose
  that $R_B$ is an analytic function defined in a complex neighborhood
  of an interval $[\al,\be]$ except at a finite number of points
  $\{\mu_1,\dots,\mu_N\}$, that the values of $R_B$ are in $B
  (L_{2,b},L_{2,-b})$ and that
  \begin{equation*}
    R_B (\la) \ph = (B - \la)^{-1} \ph \quad \text{if } \im \la > 0, \ph
  \in L_{2, b}.
  \end{equation*}
  Then, the spectrum of $B$ in the set $[\al,\be]\setminus\{\mu_1,
  \dots, \mu_N \}$ is absolutely conti\-nuous. If $\La \subset [\al,
  \be]$, $\mes\La = 0$ and $\mu_j \not\in \La$, $j = 1, \dots, N$,
  then $E_B (\La) = 0$, where $E_B$ is the spectral projector of $B$.
\end{lemma}
\noindent{\it Proof of Theorem \ref{main}.}  By Theorem \ref{RH}, the
set of all points $(k, \la) \in \RR^{d+1}$ satisfying (\ref{2}) can be
represented as the following union
\begin{equation}
  \label{eq:9}
  \{(k,\lambda)\in\R^{d+1}\text{ s.t. (\ref{2}) be satisfied}\}  
  =\bigcup_{j=1}^\infty B_{\er_j} (k_j) \times B_{\er_j} (\la_j),
\end{equation}
where, for every $j$, there exists
\begin{itemize}
\item a number $a_j \ge 0$,
\item an analytic scalar function $h_j$ defined in a complex
  neighborhood of $\overline{B_{\er_j} (k_j) \times B_{\er_j} (\la_j)}$ with
  the property
  \begin{equation*}
    \forall \la \in B_{\er_j} (\la_j) \quad \exists k \in B_{\er_j}
    (k_j) \quad \text{such that} \quad h_j (k, \la) \not= 0,
\end{equation*}
\item an analytic $B (L_{2, a_j}, L_{2, -a_j})$-valued function
  $R_H^{(j)}$ defined on the set where $h_j (k, \la) \not= 0$ and
  satisfying (\ref{17}).
\end{itemize}
Now, pick $\La \subset \RR$ such that $\mes \La = 0$. Set
\begin{gather*}
  K_0 = \{ k \in [0, 1]^d : (k+n)^2 = \la
  \text{ for some } n \in \ZZ^d, \la \in \La \},\\
  K_1 = \{ k \in [0, 1]^d : h_j (k, \la) = 0 \text{ for some } j \in
  \NN, \la \in \La \}.
\end{gather*}
Thanks to Theorem \ref{func}, we know
\begin{equation}
  \label{21}
  \mes K_0 = \mes K_1 = 0.
\end{equation}
For $k\not\in K_0$, denote
\begin{equation*}
  \La_j(k)=\{\la\in\La : (k,\la)\in B_{\er_j}(k_j)\times B_{\er_j}(\la_j)\}.
\end{equation*}
It is clear that $\La_j (k) \subset (\la_j - \er_j, \la_j + \er_j)$,
$\mes \La_j(k) = 0$, and, by~\eqref{eq:9},
\begin{equation}
  \label{eq:10}
  \La = \bigcup_{j=1}^\infty \La_j (k) \quad \forall k\not\in K_0.
\end{equation}
If $k\not\in(K_0\cup K_1)$ and $\La_j(k)\ne\emptyset$ then $h_j
(k,\la)\ne0$ for $\la\in\La_j(k)$ and $\lambda\mapsto h_j(k,\la)$ has
at most a finite number of zeros in $[\la_j-\er_j,\la_j+\er_j]$.  So
we can apply Lemma~\ref{7.1}; therefore,
\begin{equation*}
  E_{H(k)} (\La_j (k)) = 0 \quad \forall j.
\end{equation*}
This and~\eqref{eq:10} implies that
\begin{equation*}
  E_{H(k)} (\La) = 0.
\end{equation*}
Finally, one computes
\begin{equation*}
  E_H (\La) = \int_{[0, 1]^d} E_{H(k)} (\La)\,dk = \int_{[0, 1]^d
  \setminus K_0 \setminus K_1} E_{H(k)} (\La)\,dk = 0
\end{equation*}
by virtue of (\ref{21}). So, we proved that the spectral resolution of
$H$ vanishes on any set of Lebesgue measure 0, which means, by
definition, that the spectrum of the operator $H$ is purely absolutely
continuous.


\begin{thebibliography}{}
  
\bibitem{BSu} M.~Sh.~Birman, T.~A.~Suslina, {\it Periodic magnetic
    Hamiltonian with variable metrics.  Problem of absolute
    continuity}, Algebra i Analiz, vol. 11 (1999), 2, pp. 1--40
  (Russian).  English translation in St. Petersburg Math. J.  11
  (2000), no. 2, pp. 203--232.
  
\bibitem{MR1968291} A.~Boutet de Monvel, P.~Stollmann, {\it Dynamical
    localization for continuum random surface models}, Arch. Math., 80
  (2003), pp. 87--97.
  
\bibitem{MR2001b:81026} A.~Chahrour, J.~Sahbani, {\it On the spectral
    and scattering theory of the Schr{\"o}dinger operator with surface
    potential}, Rev. Math. Phys., 12 (2000), pp. 561--573.

\bibitem{E} D.~M.~Eidus, {\it Some boundary value problems in infinite
    domains}, Izvestiya AN SSSR 27 (1963), 5, pp. 1055--1080 (Russian).

\bibitem{FrSh} R.~Frank, R.~Shterenberg, {\it On the scattering theory
    of the Laplacian with a periodic boundary condition. Additional
    channels of scattering}, Doc. Math., Vol. 9 (2004), pp. 57--77.
  
\bibitem{Ge} C.~G{\'e}rard, {\it Resonance theory in atomic-surface
    scattering}, Comm. Math. Phys., 126 (1989), pp. 263--290.
  
\bibitem{MR2001m:47143} V.~Jak{\v{s}}i{\'c}, Y.~Last, {\it Corrugated
    surfaces and a.c. spectrum}, Rev. Math. Phys., 12 (2000), pp.
  1465--1503.

\bibitem{Ka:80} T.~Kato.  \newblock {\em Perturbation {Theory} for
    {Linear} {Operators}}.  \newblock Springer Verlag, Berlin, 1980.
  
\bibitem{K} P.~Kuchment, {\it Floquet theory for partial differential
    equations}, Birkh\" auser, Basel (1993).
  
\bibitem{La} O.~A.~Ladyzhenskaya, {\it On the principle of limit
    amplitude}, Uspekhi Mat. Nauk, XII N3, 75 (1957), pp. 161--164
  (Russian).
  
\bibitem{Re-Si:80} M.~Reed and B.~Simon.  \newblock {\em {Methods of
      Modern Mathematical Physics}, {Vol I}: {F}unctional {A}nalysis}.
  \newblock Academic Press, New-York, 1980.
  
\bibitem{Ya} D.~R.~Yafaev, {\it Mathematical scattering theory},
  Providence, Rhode Island, AMS, 1992.
  
\bibitem{Zw} M.~Zworski, {\it Quantum resonances and partial
    differential equations}, Proceedings of the ICM, Beijing 2002,
  vol. 3, pp. 243--254.

\end{thebibliography}

\end{document}